\documentclass[runningheads]{llncs}
\usepackage[T1]{fontenc}
\usepackage{graphicx}

\usepackage[utf8]{inputenc}
\usepackage{amsmath,cite,url}
\usepackage{algorithm}
\usepackage{algpseudocode}
\usepackage{float}
\usepackage{placeins}
\usepackage{lipsum} 
\usepackage{subfigure}
\usepackage{hyperref}
\usepackage{color}

\makeatletter
\def\blfootnote{\xdef\@thefnmark{}\@footnotetext}
\makeatother
\usepackage{fancyhdr}
\providecommand{\runtitle}{Undefined}
\providecommand{\runauthor}{Undefined}
\renewcommand{\runtitle}{Tonal Tension Conditioning in Symbolic Music Generation}
\renewcommand{\runauthor}{M. Ebrahimzadeh et al.}
\begin{document}
\title{Explicit Tonal Tension Conditioning via Dual-Level Beam Search for Symbolic Music Generation}
\author{Maral Ebrahimzadeh\inst{1} \and
Gilberto Bernardes\inst{2} \and
Sebastian Stober\inst{1}}
\institute{
Artificial Intelligence Lab, Otto-von-Guericke-University, Magdeburg, Germany
\and
University of Porto, Faculty of Engineering and INESC TEC, Porto, Portugal\\
\email{maral.ebrahimzadeh@ovgu.de}
}
\maketitle              
%
\blfootnote{\includegraphics[scale=0.25]{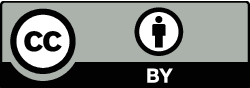} All rights remain with the authors under the Creative Commons Attribution 4.0 International License (CC BY 4.0).\\
Proc. of the 17th Int. Symposium on Computer Music Multidisciplinary Research,\\
London, United Kingdom, 2025}
\begin{abstract}
State-of-the-art symbolic music generation models have recently achieved remarkable output quality, yet explicit control over compositional features, such as tonal tension, remains challenging. We propose a novel approach that integrates a computational tonal tension model, based on tonal interval vector analysis, into a Transformer framework. Our method employs a two-level beam search strategy during inference. At the token level, generated candidates are re-ranked using model probability and diversity metrics to maintain overall quality. At the bar level, a tension-based re-ranking is applied to ensure that the generated music aligns with a desired tension curve. Objective evaluations indicate that our approach effectively modulates tonal tension, and subjective listening tests confirm that the system produces outputs that align with the target tension. These results demonstrate that explicit tension conditioning through a dual-level beam search provides a powerful and intuitive tool to guide AI-generated music. Furthermore, our experiments demonstrate that our method can generate multiple distinct musical interpretations under the same tension condition.

\end{abstract}
\section{Introduction}\label{sec:introduction}
Automatic music generation has fascinated researchers for centuries, from early stochastic experiments such as musical dice games~\cite{nierhaus2009algorithmic} to modern methods based on deep learning. Recently, Transformer models~\cite{vaswani2017attention} have shown impressive musical coherence, mirroring successes in natural language processing~\cite{deng2018deep} and computer vision~\cite{voulodimos2018deep}.

Despite these advances, the issue of interactive control remains paramount. Musicians and composers require intuitive tools to effectively guide generative models, ensuring that the output reflects their unique artistic intentions. Previous works have explored controllability over style, instrumentation, and emotion~\cite{von2023figaro,tian2025xmusic, makris2021generating,ferreira2022controlling}, yet tonal tension, a key factor that influences perceived movement and expression in melody and harmony~\cite{lerdahl2007modeling}, is significantly underexplored.

Several frameworks exist for modeling tonal tension, making it a viable direction for controllable music generation. Lerdahl’s influential model~\cite{lerdahl2007modeling} is often impractical due to manual hierarchies and parameter tuning. MorpheuS~\cite{herremans2019towards,herremans2016tension} uses Spiral Array-based metrics (dissonance, momentum, tensile strain) but lacks sensitivity to hierarchical harmony.
The Tonal Interval Vector (TIV) method~\cite{bernardes2016multi,navarro2020computational} computes the tonal tension in the Tonal Interval Space (TIS) using a discrete Fourier transform of the chroma vector. It yields a computationally efficient and perceptually relevant metric that captures multilevel pitch configurations and chord similarity, addressing the limitations of earlier models.

To achieve controllability in music generation through tonal tension, previous studies primarily employed MorpheuS metrics (cloud diameter and tensile strain)~\cite{guo2020variational,guo2022musiac,cui2024moodloopgp}, offering limited flexibility. The TIV-based method, despite its considerable potential, has been explored only once in fixed-voicing chord progression generation tasks (i.e., where the same chord cardinality is maintained throughout the progression), matching target tension curves through bio-inspired optimization techniques~\cite{navarro2020assistive}. However, reliance on evolutionary search methods severely restricts real-time controllability and seamless integration with contemporary sequence generation models. Therefore, substantial unexplored potential remains for utilizing TIV-based tonal tension metrics.

Controllability in music generation can be achieved through training-based or inference-time methods. Training-based approaches incorporate explicit control tokens for attributes like tempo or style~\cite{sun2022diffusion,guo2020variational,von2023figaro}, but require retraining and offer limited flexibility. In contrast, inference-time techniques allow dynamic plug-and-play control without modifying the underlying model~\cite{ferreira2020computer}. Beam search~\cite{poole2010artificial} can also support controllability by selecting candidate outputs based on external constraints, even when those constraints are non-differentiable.

Motivated by these limitations and opportunities, we propose to integrate a TIV-based method into Transformer-based symbolic music generation. Implementation of this approach so far was limited to voice-leading calculations for chords with identical cardinality. We improve the implementation to handle chords of varying cardinalities, reflecting realistic musical scenarios.

Our main contribution is a novel dual-level beam search strategy that explicitly guides music generation in two complementary stages. First, candidates generated at the token level are re-ranked based on model probability and diversity metrics, ensuring musical quality. Second, at the bar level, we apply tension-based re-ranking rules to align the generated output with a user-specified tonal tension curve. Our comprehensive evaluations confirm the effectiveness of the approach in both objective measures and subjective listening tests. The implementation of the dual-level beam search and tonal tension computation described in this paper is available at \url{https://github.com/MaraalE/tension-beamsearch}.

\section{Related Work}\label{sec:related_work}

\subsection{Controlling Tonal Tension in Symbolic Music Generation}
Prior works controlling tonal tension largely relied on Spiral Array metrics, as seen in MorpheuS~\cite{herremans2019towards,herremans2016tension}, which morphs music fragments via variable neighborhood search. Other works integrated Spiral Array metrics into VAEs~\cite{guo2020variational}, Transformer-based infilling models~\cite{guo2022musiac}, or LSTM networks~\cite{verstraelen2019generating}, generally emphasizing stylistic consistency over explicit tension shaping. MoodLoopGP~\cite{cui2024moodloopgp} similarly employed Spiral Array metrics, primarily targeting emotion-driven music generation. Earlier work by Melo and Wiggins~\cite{melo2003connectionist} demonstrated a neural-network-based approach to chord progression generation driven by tension curves.

While these methods often overlook critical dimensions like voice-leading, the TIV-based approach models voice-leading parsimony (i.e., minimizing motion between chords) and harmonic relationships~\cite{navarro2020computational}, though it has so far seen limited use, only in offline fixed-voicing chord generation via bio-inspired optimization techniques~\cite{navarro2020assistive}. Our work addresses this gap, integrating TIV metrics with modern Transformer-based models through dual-level beam search.

\subsection{Inference-Time Controllability in Symbolic Music Generation}
While training-based methods embed control directly into model parameters, inference-time methods dynamically guide generation without retraining. Approaches like Bardo Composer~\cite{ferreira2020computer} employ bi-objective beam search for real-time emotional control, whereas others use Monte Carlo Tree Search (PUCT)~\cite{ferreira2022controlling}, hierarchical Transformers with DTW-based re-ranking~\cite{dai2021controllable}, or text-prompt conditioned generation~\cite{atassi2023musical}. Bjare et al.~\cite{DBLP:conf/ismir/BjareLW24} recently guided music generation toward surprisal profiles using beam search based on instantaneous information content.

Most existing methods employ single-stage beam search or sampling-based adjustments. In contrast, our proposed dual-level beam search explicitly incorporates control at the token level (probability and diversity) and bar level (tonal tension alignment); This structured design facilitates more fine-grained control of musical properties.

\section{Method}\label{sec:method}
Our proposed method integrates a TIV-based tension metric into a Transformer-based music generation framework. Rather than only applying explicit tension conditioning during the training phase, our approach focuses mainly on inference-time control, enabling flexible and real-time adjustment of tonal tension without retraining the model.

\subsection{Tonal Tension Metric}
To model tonal tension, we employ a computational method based on the TIV representation~\cite{bernardes2016multi,navarro2020computational}. A fundamental building block of this model is the chroma vector, a twelve-dimensional vector representing pitch classes, commonly used to encode musical information across multiple pitch configurations such as individual pitches, chords, and keys. However, standard distance metrics applied directly to chroma vectors do not capture musically meaningful harmonic relationships. To overcome this, the TIV model applies the Discrete Fourier Transform (DFT) to chroma vectors, projecting them into a six-dimensional complex-valued space, termed the Tonal Interval Space (TIS), where intervallic structures become more distinguishable due to the DFT highlighting periodicities in pitch-class content~\cite{bernardes2016multi}.
Given two tonal interval vectors, $\mathbf{T}_1$ and $\mathbf{T}_2$, representing chords or keys projected into the TIS, we compute their distance based on musical context using two core metrics.
The \textit{Euclidean distance} between two tonal interval vectors captures perceptual similarity and consonance; smaller distances indicate smoother harmonic transitions at the same musical level (e.g., two chords that are close together tend to preserve their harmonic role in functional harmony):

\begin{equation}\label{eq:euclidean_distance}
d(\mathbf{T}_1, \mathbf{T}_2) = \sqrt{\sum_{i=0}^{N-1} \left( T_{1,i} - T_{2,i} \right)^2}
\end{equation}
In contrast, the \textit{angle} between vectors captures harmonic alignment across different levels (e.g., chord-to-key), by assessing common tone retention:
\begin{equation}\label{eq:distance_formula}
d(\mathbf{T}_1, \mathbf{T}_2) = \arccos\left(\frac{\langle \mathbf{T}_1, \mathbf{T}_2 \rangle}{\|\mathbf{T}_1\|\,\|\mathbf{T}_2\|}\right)
\end{equation}

We use three main components from a tonal tension model~\cite{navarro2020computational}, which is based on these distance metrics.

\begin{enumerate}
\item \textbf{Tonal Distance}: This includes three distinct subcomponents:
(a) Distance from the current chord to the previous chord (same musical level, thus using the Euclidean distance).
(b) Distance from the current chord to the overall key (across musical levels, thus using angle).
(c) Distance from the current chord to its tonal function (tonic I, subdominant IV, or dominant V; also angle-based).

\item \textbf{Tonal Dissonance}: It is computed as the normalized Euclidean norm subtracted from unity, aligning higher values with greater internal dissonance.
\begin{equation}
d_{\text{diss}} = 1 - \frac{\|\mathbf{T}_i\|}{\|\mathbf{T}_{\text{max}}\|}
\end{equation}
Here, $\mathbf{T}_i$ denotes the tonal interval vector of the current chord, and $\|\mathbf{T}_i\|$ its Euclidean norm. $\|\mathbf{T}_{\text{max}}\|$ is the maximum observed TIV norm across all chords, used to normalize the dissonance value between 0 and 1.

\item \textbf{Voice Leading}: This is captured by evaluating melodic stability between notes of consecutive chords. The voice-leading tension for each chord $T$ in progression $P$ is computed as:

\begin{equation}\label{eq:voice_leading}
m(T_i, P) = \sum_{l=1}^{V} e^{\frac{1}{0.05 s \mu(n_{l_{i}}, n_{l_{i-1}})}}
\end{equation}

Here, \( V \) represents the number of voices (notes) in the chord. For each voice \( l \), the term \( s \) denotes the melodic interval between consecutive notes $n_{l-{i}}$ and $n_{l_{i-1}}$, measured in semitones. The term \( \mu(n_{l_{i}}, n_{l_{i-1}}) \) measures the perceptual distance between the notes $n_{l_{i}}$ (note $l$ in chord $T_i$) and its corresponding voice $n_{l_{i-1}}$ in the previous chord $T_{i-1}$, calculated in the TIS. Thus, smaller melodic intervals and lower perceptual distances result in reduced voice-leading tension, aligning with established perceptual principles of musical stability.
Notably, the original implementation assumes chords with equal numbers of pitch classes, limiting its applicability to realistic musical contexts. To address this limitation, we use a publicly available implementation by Dmitri Tymoczko\footnote{\href{https://dmitri.mycpanel.princeton.edu/voiceleading\_utilities.py}{https://dmitri.mycpanel.princeton.edu/voiceleading\_utilities.py}} to compute the melodic interval $s$ (number of semitones), which is capable of handling chords with differing cardinalities. This modification enables more flexible and realistic voice-leading computation without altering the theoretical foundation of tonal tension.
\end{enumerate}

We compute a weighted combination of these components using scalar weights adopted from the original implementation~\cite{navarro2020computational}, where dissonance and voice leading are scaled by 30.3 and 2.71 respectively.

\subsection{Training Phase}
We adopt the REMI+\cite{von2023figaro} representation, an extension of the Revamped MIDI (REMI)\cite{huang2020pop}. REMI represents each note through four consecutive tokens encoding position, pitch, velocity, and duration, and additionally includes chord and tempo tokens. REMI+ further extends this representation by adding bar-level time-signature tokens and per-note instrument tokens, supporting multi-instrument compositions with variable time signatures. 

We use a standard Transformer model in a translation-style setup. Specifically, during training, our Transformer encoder receives bar-level control tokens, while the decoder outputs the corresponding REMI+ token sequences.
Following the setup in Figaro~\cite{von2023figaro}, we adopt control tokens for time signature, instrument list, and note density, to which we add tonal tension as an additional conditioning feature.
These attributes are encoded as discrete tokens selected from a predefined dictionary, clearly defining each bar’s musical context for the Transformer to learn from.
For instance, the pitch of a note might be represented by a token \texttt{PITCH\_32}, which is then mapped to its corresponding index in the vocabulary. The model predicts each REMI+ token sequentially conditioned on previous tokens and the provided control tokens.

\subsection{Generation Phase: Dual-Level Beam Search}
Our core contribution is a dual-level beam search strategy applied at inference (Algorithm~\ref{alg:dual_beam}), simultaneously enforcing local musical quality and diversity at the token level, and global tonal tension alignment at the bar level.

\begin{algorithm}[h]
\caption{Dual-Level Beam Search (Inference)}
\label{alg:dual_beam}
\begin{algorithmic}[1]
\Procedure{DualLevelBeamSearch}{$\mathcal{M}, \mathcal{R}, \mathcal{T}, K$}
    \State \textbf{Input:} Transformer model $\mathcal{M}$, reference piece $\mathcal{R}$, target tension curve $\mathcal{T}$, beam width $K$
    \State beams $\gets$ \{(BOS, 0)\} \Comment{Initialize beam with start token}
    \While{generation incomplete}
        \State candidates $\gets$ \{\}
        \ForAll{(seq, score) in beams}
            \State tokens $\gets$ \textsc{NucleusSample}($\mathcal{M}$, seq, $p=0.9$, $K$)
            \ForAll{token in tokens}
                \State new\_seq $\gets$ append(seq, token)
                \State bar\_cand $\gets$ current bar segment of new\_seq
                \State bar\_ref $\gets$ corresponding bar from $\mathcal{R}$
                \State div $\gets$ \textsc{DiversityMetric}(bar\_cand, bar\_ref)
                \State token\_score $\gets \text{LM}_{\text{norm}}(\text{new\_seq}) + \lambda \cdot \text{div}$
                \State candidates.add((new\_seq, token\_score))
            \EndFor
        \EndFor
        \State beams $\gets$ \textsc{TopK}(candidates, $K$)
        \If{any beam completes a bar}
            \ForAll{(seq, score) in beams}
                \State $t_{\text{cand}} \gets$ tensions of completed bars in seq
                \State $t_{\text{target}} \gets$ corresponding values from $\mathcal{T}$
                \State $t_{\text{score}} \gets$ \textsc{TensionSimilarity}($t_{\text{cand}}$, $t_{\text{target}}$)
                \State score $\gets \text{LM}_{\text{norm}}(\text{seq}) + \textit{tension\_weight} \cdot t_{\text{score}}$
            \EndFor
            \State beams $\gets$ \textsc{TopK}(beams, $K$)
        \EndIf
    \EndWhile
    \State \Return Top $K$ sequences in beams
\EndProcedure
\end{algorithmic}
\end{algorithm}

\subsubsection*{Token-Level Beam Expansion (Algorithm~\ref{alg:dual_beam}, lines 6–16)}
At each token generation step, we use nucleus (top-$p$) sampling~\cite{holtzman2019curious} to sample $K$ candidate tokens from the current Transformer context. For each active beam (partial sequence), these candidate tokens are appended to form new sequences.
Each candidate sequence is scored by combining the length-normalized log-probability of the Transformer, 
denoted $\text{LM}_{\text{norm}}$, with a diversity penalty derived from its current bar segment:

\begin{equation}
\text{Score}(x) = \text{LM}_{\text{norm}}(x) + \lambda \cdot (\text{PV}(x) + \text{DV}(x) + \text{PE}(x))
\end{equation}

where $\lambda$ is a weighting factor (0.7 in our experiments), and PV, DV, and PE represent pitch variety, duration variety, and pitch 3-gram entropy. We compare the diversity metrics of each candidate bar segment with a corresponding bar from a reference piece $\mathcal{R}$, encouraging stylistically coherent diversity. After scoring, we prune the beam, retaining the top $K$ candidates.

\subsubsection*{Bar-Level Re-ranking (Algorithm~\ref{alg:dual_beam}, lines 17–26)}
Whenever a candidate sequence completes a bar, a bar-level re-ranking step is applied. For each candidate, we calculate the average tonal tension for all completed bars, forming a candidate tension curve ($t_{\text{cand}}$), which is compared with the corresponding portion of a user-specified target tension curve $\mathcal{T}$ ($t_{\text{target}}$).

Similarity between the two tension curves is computed using Pearson correlation when the variance of $t_{\text{target}}$ exceeds a threshold ($0.001$); otherwise, absolute difference is used to ensure stability, as correlation becomes unreliable on near-flat curves. Candidates are then re-ranked using a combined score that balances the same length-normalized 
language-model score, $\text{LM}_{\text{norm}}$, with the tension similarity $t_{\text{similarity}}$:

\begin{equation}
\text{Final Score} = \text{LM}_{\text{norm}} + \textit{tension\_weight} \cdot t_{\text{similarity}}
\end{equation}

We again prune beams to retain the top $K$ sequences based on the updated final scores. Upon completion, the top-ranked sequence is considered the primary output, while remaining candidates provide alternative musical realizations under the same tension profile, highlighting our method's flexibility.

\section{Experimental Setup}\label{sec:experimental_setup}
\subsection{Dataset}
We utilize the Lakh MIDI-Matched dataset\cite{raffel2016phd}, a large-scale collection of symbolic MIDI files. Since our approach requires chord information to compute tonal tension, we preprocess all MIDI files using Midi Miner\cite{guo2019midi} to automatically identify and extract tracks containing chords. After filtering out MIDI files without chords, we obtain a final dataset of 25,555 MIDI files suitable for computing bar-level tension tokens, which we then split into training, validation, and test sets at proportions of 0.85, 0.10, and 0.05, respectively. 
\subsection{Model and Training}
We use a standard Transformer architecture with a model dimension of 512, employing 12 attention heads, 4 encoder layers, and 6 decoder layers, with a maximum sequence length of 256 tokens~\cite{von2023figaro}. Models are trained using cross-entropy loss for 12 epochs over approximately 28 hours using an NVIDIA A40 GPU.
\subsection{Inference}
During inference, we employ nucleus sampling with a threshold of 0.9, combined with a dual-level beam search. We set the beam width to 8, generating 8 candidate sequences at each token step. At the token level, we re-rank these eight candidates using a diversity weight of 0.7 and model probability scores. Subsequently, at the bar-level re-ranking stage, we select the top three beam candidates based on a tension weight of 4.0, with a sampling temperature of 0.9, thus guiding generation explicitly toward the desired tension curve.

\section{Evaluation}\label{sec:evaluation}
\subsection{Objective Evaluation}
Our evaluation assesses both the musical quality and the tonal tension accuracy of generated samples under varied training and inference conditions. Experiments isolate the impact of explicitly conditioning tension during training and inference. We measure how closely the generated output aligns with our conditioning targets using \textit{Instrument F1}, \textit{Note Density Accuracy}, and \textit{Tension Correlation}, which correspond directly to our control tokens. Additionally, we measure \textit{Groove Similarity} as an indicator of rhythmic coherence, thus reflecting overall musical quality~\cite{von2023figaro}. Evaluations are conducted primarily on 8-bar samples, with additional tests on 16-bar sequences to assess scalability.

\begin{table}
\caption{Objective evaluation metrics comparing baseline Transformer models (with and without tension conditioning) against our dual-level beam search strategy. Results for 8-bar samples represent the full test set; the 16-bar result (Dual Beam 1) uses 200 randomly selected samples. ``Dual Beam 1–3'' denote the first-, second-, and third-best candidates from our inference method (not separate models).}
\label{tab:results_overview}
\centering
\resizebox{\textwidth}{!}{
\begin{tabular}{|l|l|c|c|c|c|c|}
\hline
Model & Inference & Bars & Instr. F1 & Note Density & Groove Sim & Tension Corr \\
\hline
Baseline & Normal & 8 & 0.82 & 0.88 & 0.52 & 0.16 \\
Baseline + Tension & Normal & 8 & 0.83 & 0.62 & 0.54 & 0.18 \\
Baseline + Tension & Dual Beam 1 & 8 & 0.86 & 0.85 & 0.56 & 0.50 \\
Baseline + Tension & Dual Beam 2 & 8 & 0.86 & 0.72 & 0.55 & 0.48 \\
Baseline + Tension & Dual Beam 3 & 8 & 0.86 & 0.70 & 0.56 & 0.45 \\
Baseline + Tension & Dual Beam 1 & 16 & 0.85 & 0.62 & 0.56 & 0.42 \\
\hline
\end{tabular}
}
\end{table}

We consider three experimental conditions:
\begin{itemize}
    \item \textbf{Baseline Model}: Transformer trained using Figaro expert settings~\cite{von2023figaro} with control tokens (time signature, instrument list, note density), without tension control.
    \item \textbf{Baseline + Tension}: Baseline model augmented with an additional tension control token during training.
    \item \textbf{Dual Beam Inference (Our Main Method)}: Baseline + Tension model combined with our dual-level beam search at inference, explicitly aligning generated tension with target profiles.
\end{itemize}

Table~\ref{tab:results_overview} summarizes the key evaluation results. The baseline model (without tension control) achieves a relatively low tension correlation (0.16), indicating limited alignment with the target tension curves. Adding a tension control token during training (Baseline + Tension, normal inference) only slightly improves the tension correlation (0.18), suggesting that simply adding a control token without specialized inference is insufficient.
Our proposed dual-level beam search substantially enhances the tension correlation. The top beam candidate (Dual Beam 1) achieves a significantly higher tension correlation (0.50), indicating a good alignment with the target tension profiles. Alternative beam candidates (Dual Beam 2 and Dual Beam 3) similarly demonstrate meaningful improvements (0.48 and 0.45 respectively), confirming that our inference method consistently provides multiple viable interpretations aligned closely to target tension. Furthermore, extending the generation length to 16 bars shows a slight reduction in tension correlation (0.42), highlighting the room for future improvements to maintain control over longer musical sequences.

\begin{table}
\caption{Detailed tension correlation analysis of dual-level beam candidates, showing averages with and without filtering negative and low-variance cases.}
\label{tab:tension_detailed}
\centering
\begin{tabular}{|l|c|c|c|}
\hline
\textbf{Model} & \textbf{Avg. Tension Corr} & \textbf{Avg. Corr (Filtered)} & \textbf{Median Corr} \\
\hline
Dual Beam 1 & 0.50 & 0.62 & 0.57 \\
Dual Beam 2 & 0.48 & 0.61 & 0.56 \\
Dual Beam 3 & 0.45 & 0.58 & 0.52 \\
\hline
\end{tabular}
\end{table}

Also, Table~\ref{tab:tension_detailed} examines tension correlation by filtering out negative and low-variance cases (approximately 12\% of samples), since correlation becomes less informative for nearly constant curves. Under this analysis, Dual Beam 1 achieves the highest average correlation (0.62), demonstrating strong alignment in perceptually meaningful cases. Unfiltered results still show Dual Beam 1 leading (0.50), confirming consistent tension control across diverse outputs.

Overall, these results demonstrate that our dual-level beam search inference effectively and consistently improves tonal tension control without sacrificing other aspects of musical quality such as instrument accuracy, groove similarity, and note density. In addition, generation times average 7 minutes for an 8-bar sample, confirming feasibility for practical, offline use cases despite being slower than simple sampling methods.

\subsection{Subjective Evaluation}
\begin{figure}[htbp]
    \centering
    \subfigure[Curve 1]{\includegraphics[width=0.3\textwidth]{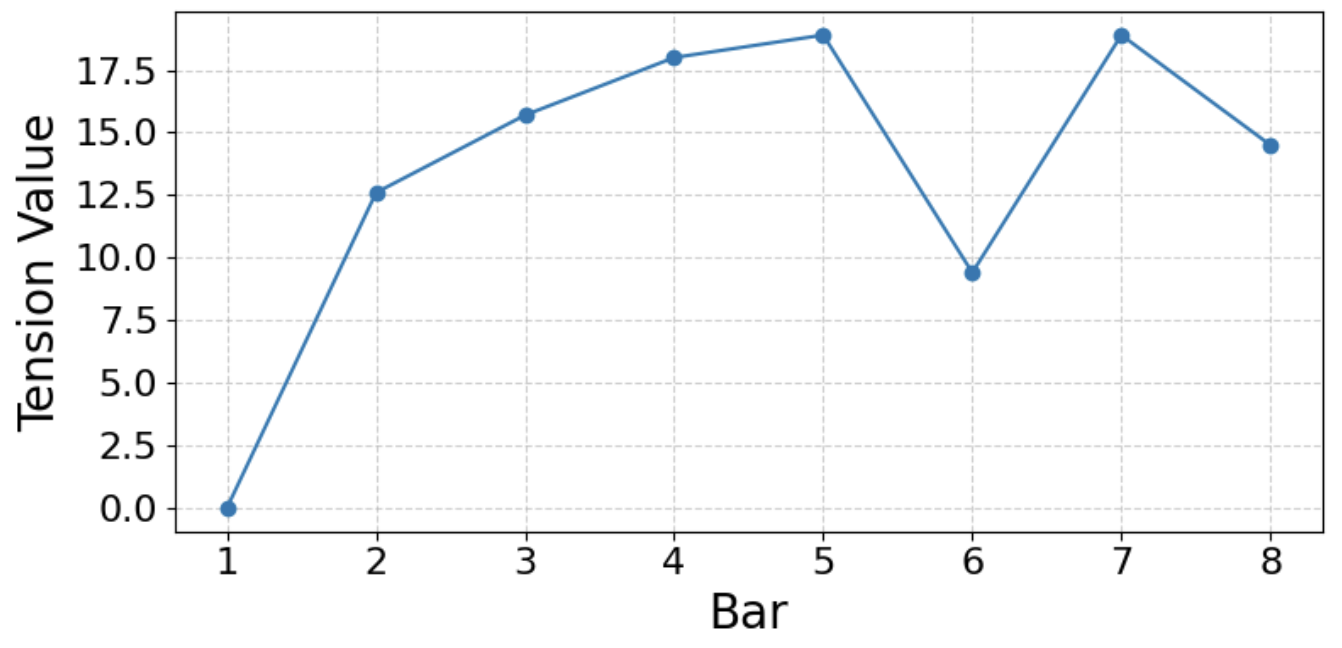}}
    \subfigure[Curve 2]{\includegraphics[width=0.3\textwidth]{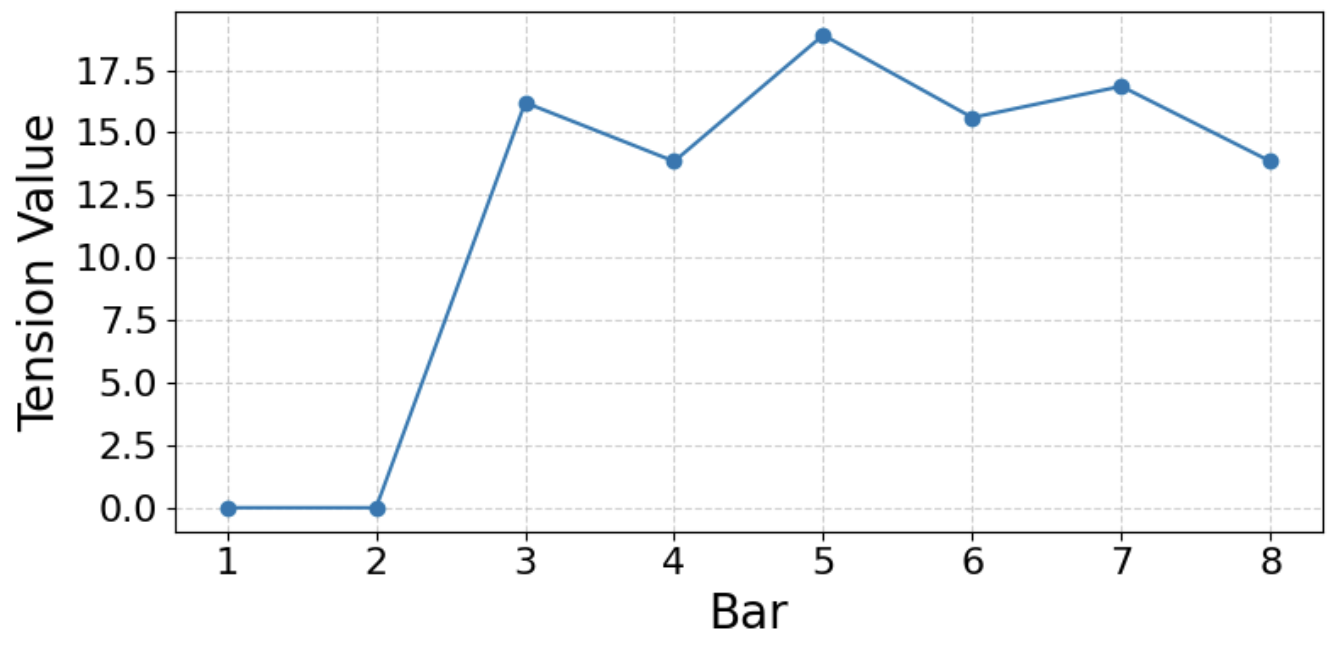}}
    \subfigure[Curve 3]{\includegraphics[width=0.3\textwidth]{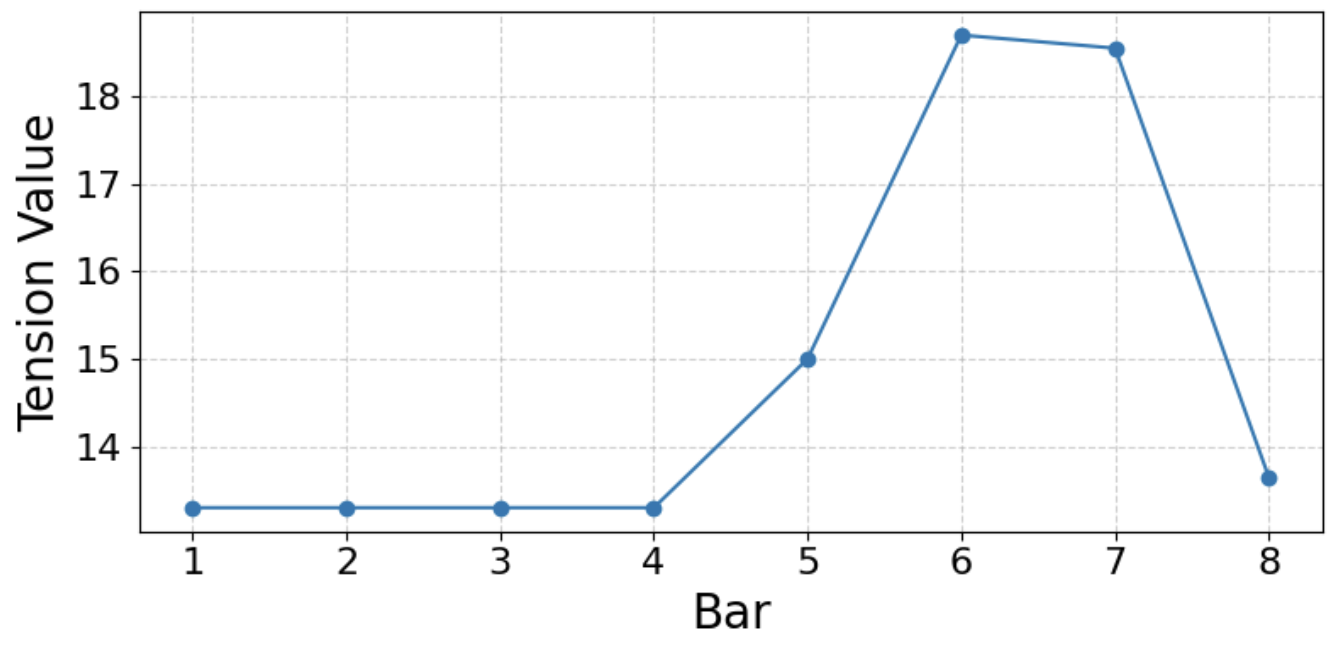}}
    
    \subfigure[Curve 4]{\includegraphics[width=0.3\textwidth]{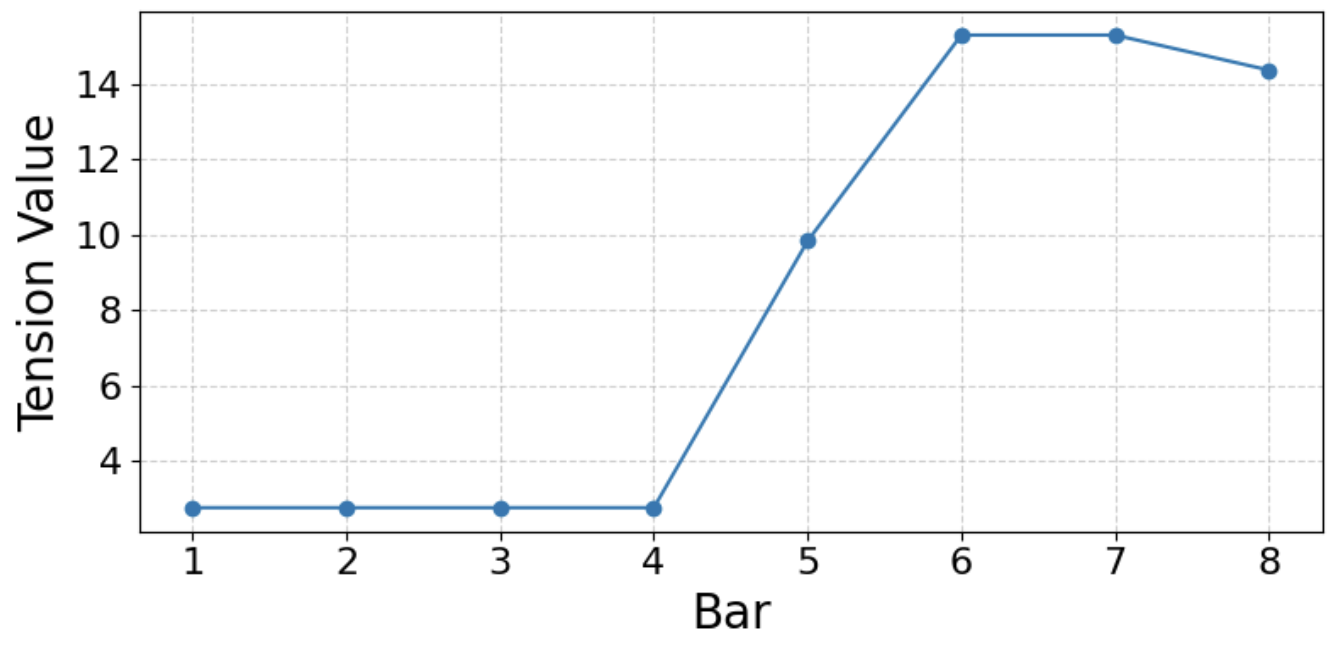}}
    \subfigure[Curve 5]{\includegraphics[width=0.3\textwidth]{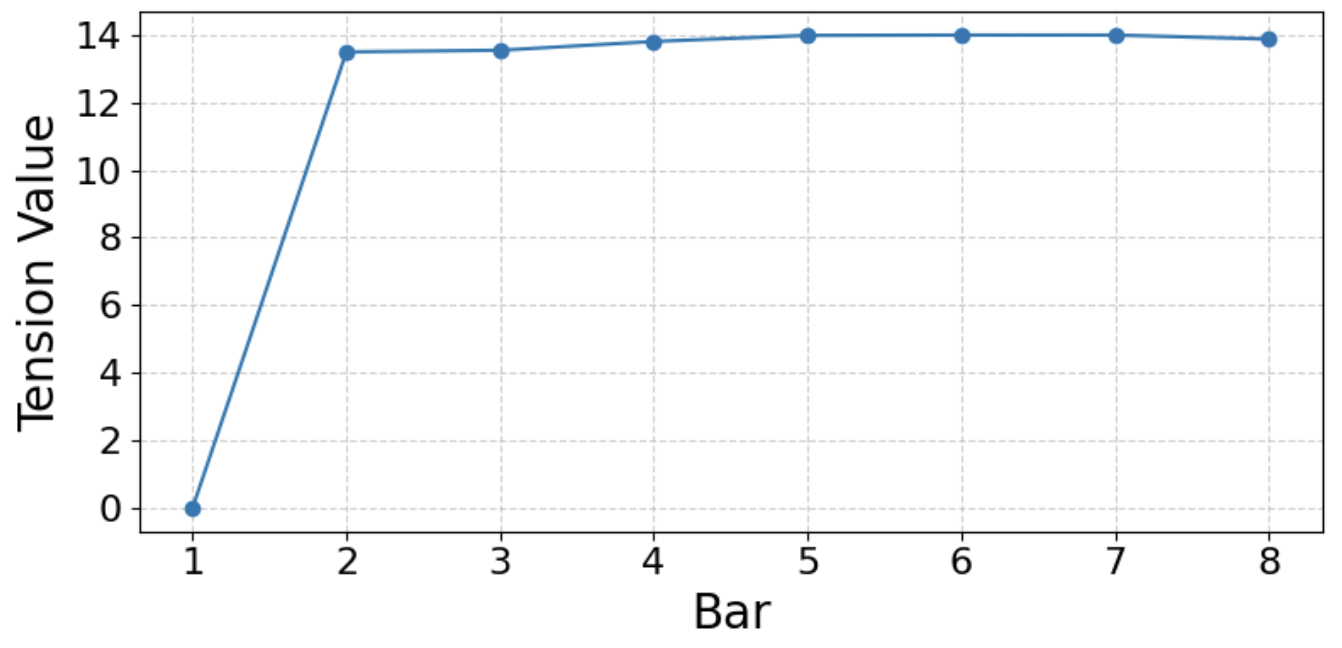}}
    
    \caption{The five tension curves used in the listening study.}
    \label{fig:tension_curves}
\end{figure}

We conducted a listening study to evaluate the perceptual effectiveness of our method specifically in terms of tonal tension alignment, which is the primary focus of this work. Five representative tension curves were selected from computed tension values of samples in the dataset (Figure~\ref{fig:tension_curves}).

\begin{figure}[h]
  \centering
  \includegraphics[width=0.6\columnwidth]{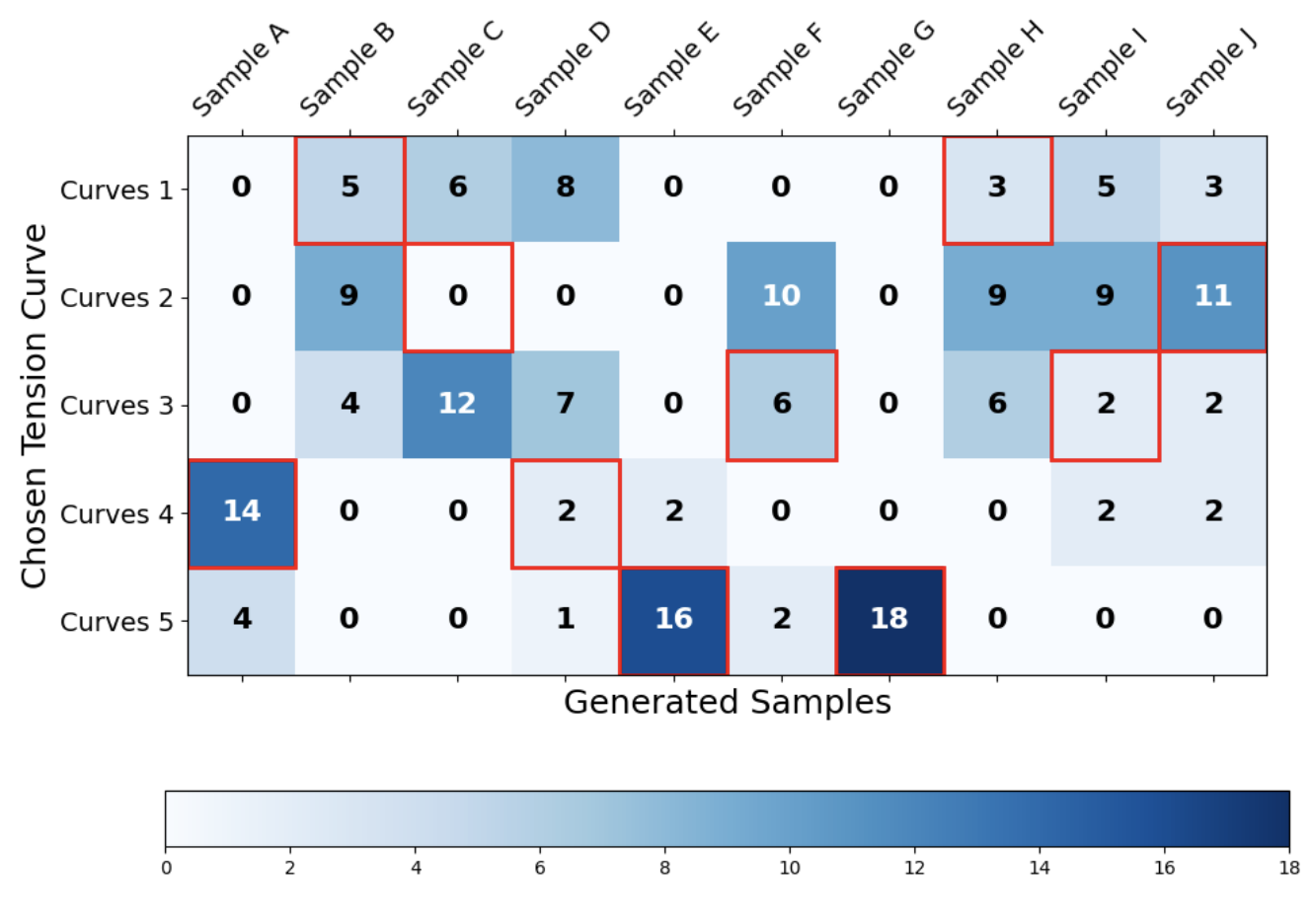} 
  \caption{Confusion matrix of listener identification of tonal tension curves (correct answers outlined in red). Curves are listed on the y-axis, and samples appear on the x-axis. Curve-to-sample assignments: Curve 1 (Samples B, H), Curve 2 (Samples C, J), Curve 3 (Samples F, I), Curve 4 (Samples A, D), Curve 5 (Samples E, G).}
  \label{fig:example}
\end{figure}

We generated two music samples per target tension curve, resulting in ten stimuli (Samples A–J). 18 participants, mostly musically experienced, were asked to identify the tension curve that best matched each sample. As shown in Figure~\ref{fig:example}, 4 out of 10 samples (A, E, G, and J) were clearly identified by most listeners, indicating strong perceptual alignment.
Many mismatches among the remaining samples were likely due to similar rising shapes, as Curves 1, 2, 3, and partially Curve 4 all exhibit upward tension profiles with variations in timing and slope. For instance, samples from Curve 1 (B, H) and Curve 3 (F, I) were often misclassified as Curve 2, suggesting listeners perceived the general rise but not finer distinctions. Considering perceptual similarity among these curves (e.g., Curves 1–3), the effective accuracy improves to 9 out of 10, indicating reliable recognition of tension trends. Curve 5 was accurately identified in both of its samples (E and G), confirming robust perceptual distinctness. Only Sample D (Curve 4) showed broader confusion, possibly due to its intermediate shape.

Overall, these findings suggest that listeners can reliably follow tension flow, particularly when curves are clearly differentiated or share similar curves. Future work may explore refining curve shapes for improved perceptual separation.
\section{Conclusion and Future Work}\label{sec:conclusion}
In this work, we introduced a dual-level beam search method for transformer-based symbolic music generation that explicitly controls tonal tension. Our method integrates tension metrics based on the tonal interval vector, effectively addressing limitations in existing tension-controlled generation models. Objective evaluations showed significant alignment improvements with target tension profiles, and subjective tests confirmed perceptual effectiveness. While subtle tension distinctions remained challenging, our approach successfully enhanced controllability without compromising musical quality.

Looking forward, our aim is to explore model interpretability through explainability methods to better understand how tonal tension conditions affect generated music. Developing interactive, real-time interfaces for composer control and extending the framework to dimensions such as emotional expressivity or rhythmic complexity, supported by perceptual studies, offer promising directions. Additionally, as the dataset exhibited key imbalance, we plan to investigate balancing techniques such as subset selection or data augmentation.

\bibliographystyle{splncs04}
\bibliography{cmmr.bib}

@book{nierhaus2009algorithmic,
  title={Algorithmic composition: paradigms of automated music generation},
  author={Nierhaus, Gerhard},
  year={2009},
  publisher={Springer Science \& Business Media}
}

@article{voulodimos2018deep,
  title={Deep Learning for Computer Vision: A Brief Review},
  author={Voulodimos, Athanasios and Doulamis, Nikolaos and Doulamis, Anastasios and Protopapadakis, Eftychios},
  journal={Computational Intelligence and Neuroscience},
  volume={2018},
  number={1},
  pages={7068349},
  year={2018},
  publisher={Wiley Online Library},
  doi={10.1155/2018/7068349}
}

@book{deng2018deep,
  title={Deep learning in natural language processing},
  author={Deng, Li and Liu, Yang},
  year={2018},
  publisher={Springer},
  doi={10.1007/978-981-10-5209-5}
}

@inproceedings{vaswani2017attention,
  title={Attention Is All You Need},
  author={Vaswani, Ashish and Shazeer, Noam and Parmar, Niki and Uszkoreit, Jakob and Jones, Llion and Gomez, Aidan N. and Kaiser, {\L}ukasz and Polosukhin, Illia},
  booktitle={Advances in Neural Information Processing Systems (NeurIPS)},
  volume={30},
  year={2017}
}

@inproceedings{von2023figaro,
  title={FIGARO: Controllable Music Generation Using Learned and Expert Features},
  author={von R{\"u}tte, Dimitri and Biggio, Luca and Kilcher, Yannic and Hofmann, Thomas},
  booktitle={The Eleventh International Conference on Learning Representations},
  year={2023}
}

@inproceedings{ferreira2022controlling,
  title={Controlling perceived emotion in symbolic music generation with monte carlo tree search},
  author={Ferreira, Lucas N and Mou, Lili and Whitehead, Jim and Lelis, Levi HS},
  booktitle={Proceedings of the AAAI Conference on Artificial Intelligence and Interactive Digital Entertainment},
  volume={18},
  number={1},
  pages={163--170},
  year={2022}
}

@inproceedings{makris2021generating,
  title={Generating lead sheets with affect: A novel conditional seq2seq framework},
  author={Makris, Dimos and Agres, Kat R and Herremans, Dorien},
  booktitle={2021 international joint conference on neural networks (IJCNN)},
  pages={1--8},
  year={2021},
  organization={IEEE}
}

@article{tian2025xmusic,
  title={XMusic: Towards a Generalized and Controllable Symbolic Music Generation Framework},
  author={Tian, Sida and Zhang, Can and Yuan, Wei and Tan, Wei and Zhu, Wenjie},
  journal={arXiv preprint arXiv:2501.08809},
  year={2025}
}

@article{lerdahl2007modeling,
  title={Modeling tonal tension},
  author={Lerdahl, Fred and Krumhansl, Carol L},
  journal={Music perception},
  volume={24},
  number={4},
  pages={329--366},
  year={2007},
  publisher={University of California Press USA}
}

@inproceedings{herremans2019towards,
  title={Towards emotion based music generation: A tonal tension modelbased on the spiral array},
  author={Herremans, Dorien and Chew, Elaine},
  booktitle={Proceedings of the Annual Meeting of the Cognitive Science Society},
  volume={41},
  year={2019}
}

@inproceedings{herremans2016tension,
  title={TENSION RIBBONS: QUANTIFYING AND VISUALISING TONAL TENSION},
  author={Herremans, Dorien and Chew, Elaine},
  booktitle={International Conference on Technologies for Music Notation and Representation--TENOR'16},
  year={2016}
}

@article{bernardes2016multi,
  title={A Multi-Level Tonal Interval Space for Modelling Pitch Relatedness and Musical Consonance},
  author={Bernardes, Gilberto and Cocharro, Diogo and Caetano, Marcelo and Guedes, Carlos and Davies, Matthew E. P.},
  journal={Journal of New Music Research},
  volume={45},
  number={4},
  pages={281--294},
  year={2016}
}

@article{navarro2020computational,
  title={A Computational Model of Tonal Tension Profile of Chord Progressions in the Tonal Interval Space},
  author={Navarro-C{\'a}ceres, Mar{\'\i}a and Caetano, Marcelo and Bernardes, Gilberto and S{\'a}nchez-Barba, Mercedes and Merch{\'a}n S{\'a}nchez-Jara, Javier},
  journal={Entropy},
  volume={22},
  number={11},
  pages={1291},
  year={2020}
}

@inproceedings{guo2020variational,
  title={A Variational Autoencoder for Music Generation Controlled by Tonal Tension},
  author={Guo, Rui and Simpson, Ivor and Magnusson, Thor and Kiefer, Chris and Herremans, Dorien},
  booktitle={Joint Conference on AI Music Creativity (CSMC \& MuMe)},
  year={2020},
  organization={AAAI Press}
}

@inproceedings{guo2022musiac,
  title={MusIAC: An extensible generative framework for music infilling applications with multi-level control},
  author={Guo, Rui and Simpson, Ivor and Kiefer, Chris and Magnusson, Thor and Herremans, Dorien},
  booktitle={International Conference on Computational Intelligence in Music, Sound, Art and Design (Part of EvoStar)},
  pages={341--356},
  year={2022},
  organization={Springer}
}

@article{navarro2020assistive,
  title={Assistive Model to Generate Chord Progressions Using Genetic Programming with Artificial Immune Properties},
  author={Navarro-C{\'a}ceres, Mar{\'\i}a and Merchan Sanchez-Jara, Javier Felix and Reis Quietinho Leithardt, Valderi and Garc{\'\i}a-Ovejero, Ra{\'u}l},
  journal={Applied Sciences},
  volume={10},
  number={17},
  pages={6039},
  year={2020},
  doi={10.3390/app10176039}
}

@inproceedings{cui2024moodloopgp,
  title={MoodLoopGP: Generating Emotion-Conditioned Loop Tablature Music with Multi-Granular Features},
  author={Cui, Wenqian and Sarmento, Pedro and Barthet, Mathieu},
  booktitle={International Conference on Computational Intelligence in Music, Sound, Art and Design (Part of EvoStar)},
  pages={97--113},
  year={2024},
  organization={Springer}
}

@article{sun2022diffusion,
  title={Diffusion-lm on symbolic music generation with controllability},
  author={Sun, Hao and Ouyang, Liwen},
  journal={ArXiv},
  year={2022}
}

@mastersthesis{verstraelen2019generating,
  title={Generating Music with Coherent Harmonic Tension},
  author={Verstraelen, Viktor},
  school={Ghent University},
  year={2019}
}

@inproceedings{ferreira2020computer,
  title={Computer-generated music for tabletop role-playing games},
  author={Ferreira, Lucas and Lelis, Levi and Whitehead, Jim},
  booktitle={Proceedings of the AAAI Conference on Artificial Intelligence and Interactive Digital Entertainment},
  volume={16},
  number={1},
  pages={59--65},
  year={2020}
}

@inproceedings{dai2021controllable,
  title={Controllable Deep Melody Generation via Hierarchical Music Structure Representation},
  author={Dai, Shuqi and Jin, Zeyu and Gomes, Celso and Dannenberg, Roger B},
  booktitle={Proceedings of the 22nd International Society for Music Information Retrieval Conference (ISMIR)},
  year={2021},
  pages={143--150},
  address={Online},
  doi={10.48550/arXiv.2109.00663}
}

@article{atassi2023musical,
  title={Musical Form Generation},
  author={Atassi, Lilac},
  journal={arXiv preprint arXiv:2310.19842},
  year={2023}
}

@book{poole2010artificial,
  title={Artificial Intelligence: foundations of computational agents},
  author={Poole, David L and Mackworth, Alan K},
  year={2010},
  publisher={Cambridge University Press}
}

@article{guo2019midi,
  title={Midi Miner--A Python library for tonal tension and track classification},
  author={Guo, Rui and Herremans, Dorien and Magnusson, Thor},
  journal={arXiv preprint arXiv:1910.02049},
  year={2019}
}

@phdthesis{raffel2016phd,
  title={Learning-based methods for comparing sequences, with applications to audio-to-MIDI alignment and matching},
  author={Raffel, Colin},
  school={Columbia University},
  year={2016}
}

@inproceedings{huang2020pop,
  title={Pop Music Transformer: Beat-Based Modeling and Generation of Expressive Pop Piano Compositions},
  author={Huang, Yu-Siang and Yang, Yi-Hsuan},
  booktitle={Proceedings of the 28th ACM international conference on multimedia},
  pages={1180--1188},
  year={2020}
}

@article{holtzman2019curious,
  title={The Curious Case of Neural Text Degeneration},
  author={Holtzman, Ari and Buys, Jan and Du, Li and Forbes, Maxwell and Choi, Yejin},
  journal={arXiv preprint arXiv:1904.09751},
  year={2019}
}

@inproceedings{DBLP:conf/ismir/BjareLW24,
author = {Mathias Rose Bjare and
Stefan Lattner and
Gerhard Widmer},
title = {Controlling Surprisal in Music Generation via Information Content
Curve Matching},
booktitle = {{ISMIR}},
pages = {922--929},
year = {2024}
}

@inproceedings{melo2003connectionist,
  title={A connectionist approach to driving chord progressions using tension},
  author={Melo, Andr{\'e}s F and Wiggins, Geraint},
  booktitle={Proceedings of the AISB Symposium on Creativity in Arts and Science},
  year={2003}
}

\end{document}